\documentstyle[epsfig,12pt]{aipproc}
\def \b{B^0}
\def \beq{\begin{equation}}
\def \eeq{\end{equation}}
\def \efi{Enrico Fermi Institute Report No. EFI}
\def \k{K^0}
\def \m{{\cal M}}
\def \ob{\overline{B^0}}
\def \ok{\overline{K^0}}
\def \s{\sqrt{2}}
\def \tl{\tilde{\lambda}}
\begin{document}
\title{CP Violation -- A Brief Review
\footnote{Invited talk presented at 2nd Tropical
Workshop in Particle Physics and Cosmology, San Juan, Puerto Rico,
May 1--6, 2000, proceedings to be published by AIP.  \efi~2000-16,
hep-ph/0005258.}}
\author{Jonathan L. Rosner}
\address{Enrico Fermi Institute and Department of Physics \\
University of Chicago, Chicago, IL 60637 USA}
\maketitle
\begin{abstract}
Some past, present, and future aspects of CP violation are reviewed.  The
discrete symmetries C, P, and T are introduced with an example drawn from
Maxwell's Equations.  The history of the discovery of CP violation in the
kaon system is described briefly, and brought up-to-date with a review of
recent results on kaon decays.  The candidate theory of CP violation, based
on phases in the Cabibbo-Kobayashi-Maskawa (CKM) matrix, will be tested
by studies of $B$ mesons, both in decays to CP eigenstates and in
``direct'' decays; we will soon learn a great deal more about whether the
CKM picture is self-consistent.  Future measurements are noted and some
brief remarks are made about the ``other'' manifestation of CP violation,
the baryon asymmetry of the Universe.
\end{abstract}

\section{Introduction}

Fundamental discrete symmetries have provided both guidance and puzzles in
our evolving understanding of elementary particle interactions.
The discrete symmetries C (charge inversion), P (parity, or space reflection),
and T (time reversal) are preserved by strong and electromagnetic
processes, but violated by weak decays.  For a brief period of several
years, it was thought that the products CP and T were preserved by all
processes, but that belief was shattered with the discovery of CP violation in
neutral kaon decays in 1964 \cite{CCFT}.  The product CPT
seems to be preserved, as is expected in local Lorentz-invariant quantum field
theories \cite{CPT}.

Since 1973 we have had a candidate theory of CP violation \cite{KM}, based
on phases in the coupling constants describing the weak charge-changing
transitions of quarks.  These couplings are described by the unitary
$3 \times 3$ {\it Cabibbo-Kobayashi-Maskawa} (CKM) \cite{KM,Cab} matrix.
This theory has survived a qualitative test with the establishment of
direct CP violation in neutral kaon decays \cite{E832,NA48}.  It is well
on its way to being tested in a wealth of $B$ decay processes.  Will these
tests be passed?  What are the implications in either case?  What will we
learn about the ``other'' manifestation of CP asymmetry in nature, the
baryon asymmetry of the Universe?  This brief review is devoted to these
questions.

In Section II we introduce the discrete symmetries P, T, and C by the
example of Maxwell's equations.  Section III is devoted to the history and
present status of CP violation and related phenomena in kaon decays,
while Section IV deals with results and prospects for $B$ mesons.  Some
future measurements are discussed in Section V.  The baryon number of the
Universe and its relation to CP violation are treated briefly in Section VI,
while Section VII concludes.

\section{Discrete symmetries}

Maxwell's equations in vacuum
provide a convenient framework for illustrating the action of discrete
symmetries, since each term in each equation must transform similarly.

Under P, we have ${\bf E}({\bf x},t) \to - {\bf E}(-{\bf x},t)$,
${\bf B}({\bf x},t) \to {\bf B}(-{\bf x},t)$,
$\nabla \to - \nabla$, ${\bf j}({\bf x},t) \to - {\bf j}({\bf -x},t)$,
i.e., electric fields change in sign while magnetic fields do not, and
currents change in direction.  Under time reversal,
${\bf E}({\bf x},t) \to {\bf E}({\bf x},-t)$,
${\bf B}({\bf x},t) \to - {\bf B}({\bf x},-t)$,
$\partial/\partial t \to - \partial/\partial t$,
${\bf j}({\bf x},t) \to - {\bf j}({\bf x}, -t)$,
i.e.,  magnetic fields change in sign while electric fields do not, since
directions of currents are reversed.  Under C,
${\bf E}({\bf x},t) \to - {\bf E}({\bf x},t)$,
${\bf B}({\bf x},t) \to - {\bf B}({\bf x},t)$,
$\rho({\bf x},t) \to - \rho({\bf x},t)$,
${\bf j}({\bf x},t) \to - {\bf j}({\bf x}, t)$,
i.e., both electric and magnetic fields change sign, since their sources
$\rho$ and ${\bf j}$ change sign.  Finally, under CPT, space and time are
inverted but electric and magnetic fields retain their signs:  ${\bf E}({\bf
x},t) \to {\bf E}({\bf -x},-t)$, ${\bf B}({\bf x},t) = {\bf B}(-{\bf x},-t)$.

The behavior of the Maxwell equations under P, T, C, and CPT is summarized in
Table 1.  Each term behaves as shown.  It is interesting that a fundamental
term in the Lagrangian behaving as ${\bf E} \cdot {\bf B}$, while Lorentz
covariant, violates P and T.  The strong suppression of such a term (as
evidenced by the small value of the neutron electric dipole moment) is known
as the {\it strong CP problem} \cite{SCPrev}, and, although of fundamental
importance, will not be discussed further here.

\renewcommand{\arraystretch}{1.4}
\begin{table}
\caption{Behavior of Maxwell's equations under discrete symmetries.}
\begin{center}
\begin{tabular}{c c c c c}
Equation & P & T & C & CPT \\ \hline
$\nabla \cdot {\bf E} = 4 \pi \rho$ & $+$ & $+$ & $-$ & $-$ \\
$\nabla \cdot {\bf B} = 0$          & $-$ & $-$ & $-$ & $-$ \\
$\nabla \times {\bf B} - \frac{1}{c} \frac{\partial {\bf E}}{\partial t} =
\frac{4 \pi}{c}{\bf j}$                 & $-$ & $-$ & $-$ & $-$ \\
$\nabla \times {\bf E} + \frac{1}{c} \frac{\partial {\bf B}}{\partial t} = 0$
                                    & $+$ & $+$ & $-$ & $-$ \\ \hline
\end{tabular}
\end{center}
\end{table}

\section{CP symmetry for kaons}

\subsection{$K \to \pi \pi$ decays}

While some neutral particles (such as $\gamma$, $Z^0$, and $\pi^0$) are equal
to their antiparticles, others (such as the neutron) are not.  The $K^0$,
discovered in cosmic radiation in the late 1940's \cite{RB}, is
one such particle.  It is characterized by an additive quantum number $S = 1$,
{\it strangeness}, introduced \cite{GN} in order to explain its strong
production (which conserves strangeness) and weak decay (which does not).  The
antiparticle of $K^0$, the $\ok$, has $S = -1$.  Since strangeness is
violated in decays, one must appeal to discrete symmetries to describe the
linear combinations of $K^0$ and $\ok$  corresponding to states of
definite mass and lifetime.  These states are
\beq \label{eqn:k1k2}
K_1 = \frac{\k + \ok}{\s}~~~,~~K_2 = \frac{\k - \ok}{\s}~~~.
\eeq
The $K_1$ is permitted to decay to $\pi \pi$ and thus should be short-lived,
while the $K_2$ is forbidden to decay to $\pi \pi$, must instead decay to
$3 \pi$, $\pi \ell \nu_\ell$, etc., and thus will be longer-lived.  Indeed,
the short-lived neutral kaon ($\sim K_1$)
lives for only 0.089 ns, while the long-lived neutral kaon ($\sim K_2$)
lives for 52 ns, nearly a factor of 600 longer. 

The original argument by Gell-Mann and Pais \cite{GP}, based in 1955 on C and P
conservation, was recast in 1957 in terms of the product CP \cite{CPK}, to
correspond to the newly formulated CP-invariant theory of the weak
interactions.  The $K^0$ and $\ok$ have spin zero.  A spin-zero final state of
$\pi \pi$ has CP eigenvalue equal to $+1$.  Thus, if CP is conserved, it is the
CP-even linear combination of $K^0$ and $\ok$ which decays to $\pi \pi$.  With
a phase convention such that $CP |K^0 \rangle
= | \ok \rangle$, this is just the combination $K_1$.
The Gell-Mann--Pais proposal was soon confirmed \cite{KL} by the discovery of
the predicted long-lived particle corresponding to $K_2$.

Similar behavior is encountered in many cases of degenerate systems, such as
two coupled pendula \cite{BW} or a drum-head in its first excited state.  In
the latter case, the drum has two degenerate modes, each with one nodal line
corresponding to a diameter, which will be orthogonal to one another if the
corresponding nodal lines are perpendicular to each other.  Consider two
equally valid bases:

\begin{itemize}

\item{(B1)} Diagonal nodal lines point to the upper right ($R$)
and the upper left ($L$).

\item{(B2)} The nodal lines are horizontal ($H$) and vertical ($V$).

\end{itemize}

We can draw the analogy $R \leftrightarrow \k$, $L \leftrightarrow \ok$.
Suppose, now, that a fly alights on the bottom edge of the drum head, such
that it sits on the nodal line of the $V$ mode.  Then the modes $V$ and $H$
are split from one another.  The mode $H = (R + L)/\s$ which couples to the
fly will shift in mass and lifetime.  It is analogous to $K_1$ and the fly
is analogous to the $\pi \pi$ system.  The mode $V$ is unaffected by the
fly.  It is analogous to $K_2$.

In 1964, Christenson, Cronin, Fitch, and Turlay \cite{CCFT}, using a spark
chamber exposed to a beam of long-lived neutral kaons, found that these
particles indeed {\it did} decay to $\pi \pi$.  For many years this phenomenon
could be described in terms of a single parameter $\epsilon$, such that the
states of definite mass and lifetime become
\beq \label{eqn:mix}
K_1 \to K_S~({\rm ``short"}) \simeq K_1 + \epsilon K_2~~,~~~
K_2 \to K_L~({\rm ``long"})  \simeq K_2 + \epsilon K_1~~~,
\eeq
with $|\epsilon| \simeq 2 \times 10^{-3}$, and Arg($\epsilon) \simeq \pi/4$.
Confirmation of this description was provided by the rate asymmetry in the
decays $K_L \to \pi^\pm \ell^\mp \nu_\ell$, which measures Re $\epsilon$.
But what is the source of $\epsilon$?

One possibility was suggested almost immediately by Wolfenstein \cite{SW}:
A new ``superweak'' $|\Delta S = 2|$ interaction could mix $\k = d \bar s$ and
$\ok = s \bar d$ (where $d$ and $s$ denote quarks) without any other observable
consequences.  This theory would imply, for example, that no difference in
the ratio of CP-violating and CP-conserving amplitudes would arise when
comparing $\pi^+ \pi^-$ and $\pi^0 \pi^0$ final states.

A new opportunity for generating not only $\epsilon$ but other CP-violating
effects as well arises when there are at least three quark families, as
first proposed by Kobayashi and Maskawa \cite{KM}.  Loop diagrams inducing
the transition $d \bar s \leftrightarrow s \bar d$ involving internal lines of
$W^+ W^-$ and $u,c,t$ quarks and antiquarks can lead to $\epsilon \ne 0$ when
the coupling constants are complex.  With three quark families, one cannot
redefine phases of quarks so that all the couplings are real.  Some other
consequences of the Kobayashi-Maskawa theory will be mentioned presently.

The time-dependence of the two-component $\k$ and $\ok$ system is governed by
a $2 \times 2$ {\it mass matrix} $\m$ (for reviews see \cite{Revs}):
\beq
i \frac{\partial}{\partial t} \left[ \begin{array}{c} \k \\ \ok \end{array}
\right] = \m \left[ \begin{array}{c} \k \\ \ok \end{array} \right]~~~,
\eeq
where $\m = M - i \Gamma/2$, and $M$ and $\Gamma$ are Hermitian matrices.
The eigenstates are, approximately,
\beq
K_S \simeq K_1 + \epsilon K_2~~,~~~K_L \simeq K_2 + \epsilon K_1~~~,
\eeq
corresponding to the eigenvalues $\mu_{S,L} = m_{S,L} - i \gamma_{S,L}/2$, with
\beq
\epsilon \simeq \frac{{\rm Im}(\Gamma_{12}/2) + i~{\rm Im}~M_{12}}
{\mu_S - \mu_L}~~~.
\eeq
Using both data and the magnitude of CKM matrix elements one can show
\cite{Revs} that the second term dominates.  Since the mass difference
$m_L - m_S$ and width difference $\gamma_S - \gamma_L$ are nearly equal,
the phase of $\mu_L - \mu_S$ is about $\pi/4$, so that the phase of $\epsilon$
is also $\pi/4$ (mod $\pi$).

It is easy to emulate the {\it CP-conserving} neutral kaon system in
table-top demonstrations of systems with two degenerate states, such as the
pair of coupled pendula mentioned above \cite{BW}.  The demonstration of CP
violation is harder, requiring systems that emulate Im($M_{12}) \ne 0$ or
Im($\Gamma_{12}) \ne 0$.  One can couple two identical resonant circuits
``directionally'' to each other so that the energy fed from circuit 1 to
circuit 2 differs from that fed in the reverse direction \cite{TTTV}.  Devices
with this property utilize Faraday rotation of the plane of polarization of
radio-frequency waves.  More recently, it was realized \cite{RS} that this
asymmetric coupling is inherent in the equations of motion of a spherical
(or ``conical'') pendulum in a rotating coordinate system, giving rise to
the precession of the plane of oscillation of the Foucault pendulum.  In
either case the analogy actually deepens the mystery of CP violation, since
the CP-violating effect is imposed, so to speak, ``from the outside,''
using a magnetic field in the case of directional couplers or a rotating
coordinate frame in the case of the Foucault pendulum. 

To return to the CKM matrix, we have the following parameterization
suggested by Wolfenstein \cite{WP}:
\beq
V \equiv \left[ \begin{array}{c c c}
V_{ud} & V_{us} & V_{ub} \\
V_{cd} & V_{cs} & V_{cb} \\
V_{td} & V_{ts} & V_{tb} \\ \end{array} \right] = \left[ \begin{array}{c c c}
1 - \frac{\lambda^2}{2} & \lambda & A \lambda^3 (\rho - i \eta) \\
- \lambda & 1 - \frac{\lambda^2}{2} & A \lambda^2 \\
A \lambda^3(1 - \rho - i \eta) & - A \lambda^2 & 1 \end{array} \right]~~~,
\eeq
where $\lambda = \sin \theta_{\rm C} \simeq 0.22$ describes strange particle
decays.  Here $\theta_{\rm C}$ is the Gell-Mann--L\'evy--Cabibbo \cite{Cab,GL}
angle, originally introduced to preserve the universal strength of the
hadronic weak current.  The unitarity of the CKM matrix, $V^\dag = V^{-1}$, is
the modern way of implementing this requirement.

We learn $|V_{cb}| = A \lambda^2 \simeq 0.039 \pm 0.003$ from the dominant
decays of $b$ quarks, which are to charmed quarks \cite{CKMrevs}.  (We have
expanded errors somewhat in comparison with those quoted in some reviews
\cite{Parodi}.  The
dominant source of error in many cases is theoretical.)  Similarly, charmless
$b$ decays give $|V_{ub}/V_{cb}| = 0.090 \pm 0.025 = \lambda (\rho^2 +
\eta^2)^{1/2}$, leading to a constraint on $\rho^2 + \eta^2$.

As a result of the unitarity of the CKM matrix, the quantities $V^*_{ub}/A
\lambda^3 = \rho + i \eta$, $V_{td}/A \lambda^3 = 1 - \rho - i \eta$, and 1
form a triangle in the $(\rho,\eta)$ plane (Fig.~1).
The angles opposite these sides
are, respectively, $\beta = -{\rm Arg}(V_{td})$, $\gamma = {\rm
Arg}(V^*_{ub})$, and $\alpha = \pi - \beta - \gamma$.  We still do not have
satisfactory limits on the angle $\gamma$ (equivalently, on the magnitude of
the side $V_{td}$) of this ``unitarity triangle.''  Further information comes
from the following constraints (see \cite{JRlatt} for more details):

\begin{figure}
\centerline{\epsfysize = 1.8in \epsffile {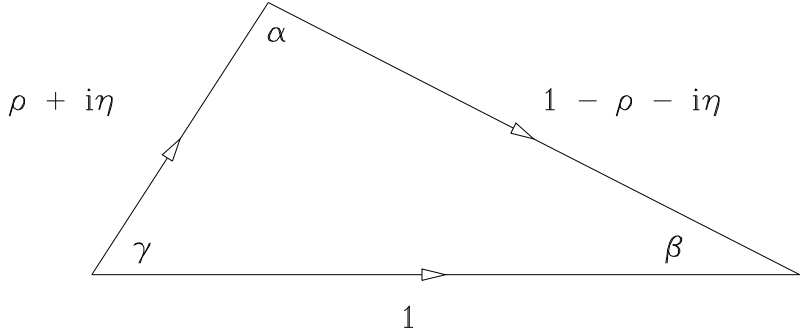}}
\caption{Unitarity triangle for CKM elements.  Here $\rho + i \eta =
V^*_{ub}/A \lambda^3$; $1 - \rho - i \eta = V_{td}/A \lambda^3$.} 
\end{figure}

\begin{enumerate}

\item The magnitude of $\epsilon$ constrains mainly the imaginary part of
$V_{td}^2$, which is proportional to
$\eta(1-\rho)$, since the top quark dominates the
loop diagram giving rise to $\k$--$\ok$ mixing.  A correction due to charmed
quarks changes the 1 to 1.44, with the result $\eta(1.44 - \rho) = 0.51 \pm
0.18$.

\item We have taken the amplitude for mixing of the neutral $\b$ meson with its
antiparticle $\ob$ to be $\Delta m_d = 0.473 \pm 0.016$ ps$^{-1}$ \cite{BOSC}.
The subscript $d$ denotes the light quark in the $\b$.  Taking
the matrix element of the four-quark operator inducing the relevant $\bar b d
\leftrightarrow \bar d b$ transition to be $f_B \sqrt{B_B} = 200 \pm 40$ MeV,
we find a constraint on $|V_{td}|$ which amounts to $|1 - \rho - i \eta| =
1.01 \pm 0.21$.

\item We have used the following lower limit for mixing of the strange $b$
meson $B_s = \bar b s$ with its antiparticle:  $\Delta m_s > 14.3$ ps$^{-1}$
(95\% c.l.) \cite{BOSC,Blay}.  Since the relevant CKM elements (including
$|V_{ts}| = A \lambda^2$) are fairly well known, this result serves mainly to
constrain the combination of hadronic parameters $f_{B_s} \sqrt{B_{B_s}}$ and
hence, through the assumption $[f_{B_s} \sqrt{B_{B_s}}]/[f_B \sqrt{B_B}] <
1.25$ \cite{JRFM}, yields the bound $|V_{ts}/V_{td}| > 4.3$ or $|1 - \rho - i
\eta| < 1.05$.

\end{enumerate}

The resulting limits on $(\rho,\eta)$ are a roughly rectangular region
bounded on the left by $|1 - \rho - i \eta| < 1.05$, on the top and bottom
by $0.3 < (\rho^2 + \eta^2)^{1/2} < 0.52$, and on the right by $|1 - \rho - i
\eta| > 0.8$.  Only a small region is excluded by the bound arising from
the parameter $\epsilon$:  $\eta(1.44 - \rho) > 0.33$.  Even without this
bound, the case of real CKM matrix elements ($\eta = 0$), i.e., a superweak
origin for $\epsilon$, is disfavored.  The boundaries of this region give rise
to the minimum and maximum values of $\alpha,\beta,\gamma$ shown in Table 2.
These bounds imply
\beq
-0.71 < \sin 2 \alpha < 0.59~~,~~~
 0.59 < \sin 2 \beta  < 0.89~~,~~~
 0.54 < \sin^2 \gamma < 1~~~
\eeq
for quantities which are measurable in $B$ decays (see below).  The allowed
values of $(\rho,\eta)$ are $\simeq (0.14 \pm 0.15,~0.38 \pm 0.13)$.

\begin{table}
\caption{Ranges of angles in the unitarity triangle.}
\begin{center}
\begin{tabular}{c c c c c c}
Angle    &       Expression        &  & Degrees & $\rho$ & $\eta$ \\ \hline
$\alpha$ & $\pi - \beta - \gamma$    & Min &  72 & $-0.01$ & 0.30 \\
         &                           & Max & 113 &  0.25   & 0.27 \\
$\beta$  & tan$^{-1}[\rho/(1-\eta)]$ & Min &  17 & $-0.01$ & 0.30 \\
         &                           & Max &  31 &  0.29   & 0.43 \\
$\gamma$ & tan$^{-1}(\eta/\rho)$     & Min &  48 &  0.25   & 0.27 \\  
         &                           & Max &  92 & $-0.01$ & 0.30 \\ \hline
\end{tabular}
\end{center}
\end{table}

The Kobayashi-Maskawa theory predicts small differences in CP-violating
decays to pairs of charged and neutral pions.  These arise in the
following way.

\begin{enumerate}

\item ``Tree'' amplitudes are
governed by $\bar s \to \bar u u \bar d$.  Since this subprocess has three
nonstrange quarks in the final state, it contributes to both $\Delta I = 1/2$
and $\Delta I = 3/2$ transitions, and hence to both $I_{\pi \pi} = 0$ and
$I_{\pi \pi} = 2$ final states.  The corresponding CKM matrix elements are
real, so these amplitudes do not have a weak phase.

\item ``Penguin'' amplitudes involve a transition $\bar s \to \bar d$ with
internal $W$ and $u,c,t$ lines and emission or absorption of a gluon.
The subprocess has only one nonstrange quark
in the final state so it contributes only to $\Delta I = 1/2$ transitions
and hence only to the $I_{\pi \pi} = 0$ final state.  Because of the presence
of all three $Q=2/3$ quarks in internal lines, these amplitudes have a weak
phase.

\end{enumerate}

As a consequence of the different isospin structure and weak phases of
the tree and penguin amplitudes, the $I_{\pi \pi} = 0$ and $I_{\pi \pi} = 2$
amplitudes thus acquire different weak phases, leading to a small difference
from unity of the ratio
\beq
R \equiv \frac{\Gamma(K_L \to \pi^+ \pi^-)/\Gamma(K_S \to \pi^+ \pi^-)}
{\Gamma(K_L \to \pi^0 \pi^0)/\Gamma(K_S \to \pi^0 \pi^0)}
= 1 + 6 {\rm~Re} \frac{\epsilon'}{\epsilon}~~~,
\eeq
where $\epsilon'$ is related to the imaginary part of the ratio of the
$I_{\pi \pi} = 2$ and $I_{\pi \pi} = 0$ amplitudes.  The ratio $\epsilon'/
\epsilon$ acquires an important term proportional to the CKM parameter $\eta$
from the penguin contribution.  This term is partially canceled by an
``electroweak penguin'' in which the gluon mentioned above is replaced by a
virtual photon or $Z$, whose isospin-dependent couplings to quarks
induce $\Delta I = 3/2$ contributions.  $\epsilon'/\epsilon$ is expected to
be nearly real.  Its magnitude was estimated by one group \cite{Buras} to be a
few parts in $10^4$, with a broad and somewhat asymmetric probability
distribution extending from slightly below zero to above $2 \times 10^{-3}$.
Some other estimates, discussed in Refs.~\cite{K99}, permit higher values.

\begin{table}
\caption{Experimental values for Re$(\epsilon'/\epsilon)$.}
\begin{center}
\begin{tabular}{c c c c}
\protect
Experiment & Reference & Value ($\times 10^{-4}$) & $\Delta \chi^2$ \\ \hline
Fermilab E731 & \cite{E731} &  $7.4 \pm 5.9$ & 3.97 \\
CERN NA31     & \cite{NA31} & $23.0 \pm 6.5$ & 0.35 \\
Fermilab E832 & \cite{E832} & $28.0 \pm 4.1$ & 4.65 \\
CERN NA48     & \cite{NA48} & $14.0 \pm 4.3$ & 1.44 \\
Average       &             & $19.2 \pm 4.6$ & $\sum = 10.4$ \\ \hline
\end{tabular}
\end{center}
\end{table}

The most recent experiments on Re($\epsilon'/\epsilon$) are summarized in
Table 3.  A scale factor \cite{PDG} of 1.86 is included in the error of the
average to account for the large spread in quoted results. The value of
$\epsilon'/\epsilon$ is non-zero, with a magnitude in the ballpark of
estimates based on the Kobayashi-Maskawa theory.  The fact that it is larger
than some theoretical estimates is not a serious problem, given that we still
cannot account reliably for the
large enhancement of $\Delta I = 1/2$ amplitudes with respect to $\Delta I =
3/2$ amplitudes in {\it CP-conserving} $K \to \pi \pi$ decays.

\subsection{Other rare kaon decays}

A CP- or T-violating angular asymmetry in $K_L \to \pi^+ \pi^- e^+ e^-$ has
recently been reported \cite{KTeVa,NA48a}.  With a final state consisting of
four distinct particles, using the three independent final c.m. momenta, one
can construct a T-odd observable whose presence is signaled by a characteristic
distribution in the angle $\phi$ between the $\pi^+ \pi^-$ and $e^+ e^-$ planes.

The asymmetry in $\sin \phi \cos \phi$ reported in Ref.~\cite{KTeVa}
is $(13.6 \pm 2.5 \pm 1.2)\%$.  It arises from interference between two
processes.  (1) The $K_L$ decays to $\pi^+ \pi^-$ with an amplitude $\epsilon$.
This process is CP-violating.  One of the pions then radiates a virtual photon
which internally converts to $e^+ e^-$.  (2) The CP-odd state $K_2$ can
decay directly to $\pi^+ \pi^- \gamma$ via a weak magnetic dipole transition.
This process is CP-conserving.

The decay $K_L \to \mu^+ \mu^- \gamma$ has recently been studied
with sufficiently high statistics to permit a greatly improved measurement
of the virtual-photon form factor in $K_L \to \gamma^* \gamma$ \cite{BQ}.
This measurement is useful in estimating the long-distance contribution to
the real part of the amplitude in $K_L \to \gamma^{(*)} \gamma^{(*)} \to
\mu^+ \mu^-$, which in turn allows one to limit the short-distance contribution
to $K_L \to \mu^+ \mu^-$.  Since this contribution involves loops with
virtual $W$'s and $u,c,t$ quarks, useful bounds on CKM matrix elements can be
placed.  Preliminary results \cite{BQ} indicate $\rho > -0.2$, the best limit
so far from any process involving kaons.

Several neutral-current processes involving $K \to \pi + ({\rm lepton
~pair})$ can shed further light on the Kobayashi-Maskawa theory of CP
violation \cite{BuK}.

\begin{enumerate}

\item The decay $K^+ \to \pi^+ \nu \bar \nu$ is sensitive primarily to
$|V_{td}|$, with a small charm correction, and so constrains the combination
$|1.4 - \rho - i \eta|$.  The predicted branching ratio is roughly
\beq
{\cal B}(K^+ \to \pi^+ \nu \bar \nu) \simeq 10^{-10} \left| \frac{1.4 - \rho
- i \eta}{1.4} \right|^2~~~,
\eeq
For $0 \le \rho \le 0.3$ one then predicts (see \cite{BuK}) ${\cal B}(K^+ \to
\pi^+ \nu \bar \nu) = (0.8 \pm 0.2) \times 10^{-10}$, with additional
uncertainties associated with the charmed quark mass and the
magnitude of $V_{cb}$.  A measurement of this branching ratio with an
accuracy of 10\% is of high priority in constraining $(\rho,\eta)$ further.

The Brookhaven E787 Collaboration has reported one event
with negligible background \cite{E787}, corresponding to
\beq
{\cal B}(K^+ \to \pi^+ \nu \bar \nu) = (1.5^{+3.4}_{-1.2}) \times 10^{-10}~~~.
\eeq
More data are expected from the final stages of analysis of this experiment,
as well as from a future version (Brookhaven E949) with improved sensitivity.

\item The decays $K_L \to \pi^0 \ell^+ \ell^-$ are expected to be dominated by
CP-violating contributions, both indirect ($\sim \epsilon$) and direct. 
There is also a CP-conserving ``contaminant'' from the intermediate state
$K_L \to \pi^0 \gamma \gamma$. The direct contribution probes the CKM parameter
$\eta$.  It is expected to be comparable in magnitude to the indirect
contribution, and to have a phase of about $\pi/4$ with respect to it.  Each
contribution (including the CP-conserving one) is expected to correspond to
a $\pi^0 e^+ e^-$ branching ratio of a few parts in $10^{12}$.
However, the decay $K_L \to \pi^0 e^+ e^-$ may be limited by backgrounds
in the $\gamma \gamma e^+ e^-$ final state associated with radiation of a
photon in $K_L \to \gamma e^+ e^-$ from one of the leptons
\cite{HG}.  Present experimental upper limits (90\% c.l.) \cite{pll} are
\beq
{\cal B}(K_L \to \pi^0 e^+ e^-) < 5.64 \times 10^{-10}~,~~
{\cal B}(K_L \to \pi^0 \mu^+ \mu^-) < 3.4 \times 10^{-10}~~,
\eeq
still significantly above theoretical expectations.

\item The decay $K_L \to \pi^0 \nu \bar \nu$ is expected to be due entirely to
CP violation, and provides a clean probe of $\eta$.  Its branching
ratio, proportional to $A^4 \eta^2$, is expected to be about $3 \times
10^{-11}$.  The best current experimental upper limit (90\% c.l.) for this
process \cite{pnn} is ${\cal B}(K_L \to \pi^0 \nu \bar \nu) < 5.9 \times
10^{-7}$, several orders of magnitude above the expected value.

\end{enumerate}

\subsection{Is the CKM picture of CP violation right?}

Two key tests have been passed so far.  The theory has succeeded, albeit
qualitatively, in predicting the range Re$(\epsilon'/\epsilon) = (1~{\rm to}~2)
\times 10^{-3}$.  Its prediction for the branching ratio for $K^+ \to \pi^+
\nu \bar \nu$ is in accord with the experimental rate deduced from the one
event observed so far.

One test still to be passed in the decays of neutral kaons is the measurement
of the height $\eta$ of the unitarity triangle through the decay $K_L \to
\pi^0 \nu \bar \nu$.  Prospects for this measurement will be mentioned below.
However, in the nearer term, one looks forward to a rich set of effects in
decays of particles containing $b$ quarks, particularly the $B$ mesons.  To
this end, experiments are under way at a number of laboratories around the
world.

Asymmetric $e^+ e^-$ collisions are being studied at two ``$B$ factories,''
the PEP-II machine at SLAC with the BaBar detector, and the KEK-B collider in
Japan with the Belle detector.  By end of April 2000, these detectors
were recording about 100 and 60 pb$^{-1}$ of data per day, respectively, and
had accumulated about 6 and 2 fb$^{-1}$ of data at the energy of the
$\Upsilon(4S)$ resonance, which decays almost exclusively to $B \bar B$.  The
BaBar experiment expects to have about 100 tagged $B^0 \to J/\psi K_S$ decays
by this coming summer \cite{SS}.

Significant further data on $e^+ e^-$ collisions at the $\Upsilon(4S)$
are expected from the Cornell Electron Storage Ring with the upgraded CLEO-III
detector.  The HERA-b experiment at DESY in Hamburg will study $b$ quark
production via the collisions of 920 GeV protons with a fixed target.  The
CDF and D0 detectors at Fermilab will devote a significant part of their
program at Run II of the Tevatron to $B$ physics.  In the longer term, one can
expect further results on $B$ physics from the general-purpose LHC detectors
ATLAS and CMS and the dedicated LHC-b detector at CERN, and possibly the
dedicated BTeV detector at Fermilab.

\section{CP violation and $B$ decays}

In constrast to the neutral kaon system, in which the eigenstates of the
mass matrix differ in lifetime by nearly a factor of 600, the eigenstates of
the corresponding $\b$--$\ob$ mass matrix are expected to differ in lifetime
by at most 10--20\% for strange $B$'s \cite{BBD}, and considerably less
for nonstrange $B$'s.  Thus, instead of studying the properties of mass
eigenstates like $K_L$, one must resort to other means.  There are two main
avenues of study.

\begin{itemize}

\item {\it Decays to CP eigenstates $f = \pm {\rm CP}(f)$} utilize interference
between direct decays $\b \to f$ or $\ob \to f$ and the corresponding paths
involving mixing:  $\b \to \ob \to f$ or $\ob \to \b \to f$.  Final states
such as $f = J/\psi K_S$ provide ``clean'' examples in which one quark
subprocess is dominant.  In this case one measures $\sin 2 \beta$ with
negligible corrections.  For the final state $\pi^+ \pi^-$, one measures
$\sin 2 \alpha$ only to the extent that the direct decay is dominated by
a ``tree'' amplitude (the quark subprocess $b \to u \bar u d$).  When
contamination from the penguin subprocess $b \to d$ is present (as it is
expected to be at the level of several tens of percent), one must measure
decays to other $\pi \pi$ states (such as $\pi^\pm \pi^0$ and $\pi^0 \pi^0$)
to sort out various decay amplitudes \cite{GrL}.

\item {\it ``Self-tagging'' decays} involve final states $f$ such as $K^+
\pi^-$ which can be distinguished from their CP-conjugates $\bar f$.  A
CP-violating rate asymmetry arises if there exist two weak amplitudes
$a_i$ with weak phases $\phi_i$ and strong phases $\delta_i$ ($i=1,2)$:
$$
A(B \to f) = a_1 e^{i(+\phi_1 + \delta_1)} + a_2 e^{i(+\phi_2 + \delta_2)}~~~,
$$
\beq
~~~~~A(\bar B \to \bar f) = a_1 e^{i(-\phi_1 + \delta_1)} + a_2 e^{i(-\phi_2
 + \delta_2)}~~~.
\eeq
Note that the weak phase changes sign under CP-conjugation, while the strong
phase does not.  The rate asymmetry is then
\beq
{\cal A}(f) \equiv \frac{\Gamma(f) - \Gamma(\bar f)}
{\Gamma(f) + \Gamma(\bar f)}
= \frac{2 a_1 a_2 \sin(\phi_1 - \phi_2) \sin(\delta_1 - \delta_2)}
{a_1^2 + a_2^2 + 2 a_1 a_2 \cos(\phi_1 - \phi_2) \cos (\delta_1 - \delta_2)}~~.
\eeq
Thus the two amplitudes must have different weak {\it and} strong phases in
order for a rate asymmetry to be observable.  The
CKM theory predicts the weak phases, but no reliable estimates of strong phases
in $B$ decays exist.  Some ways of circumventing this difficulty will be
mentioned.

\end{itemize}

\subsection{Decays to CP eigenstates}

The interference between mixing and decay in decays of neutral $B$ mesons to
CP eigenstates leads to a term which modulates the exponential decay (see,
e.g., \cite{DR}):
\beq
\frac{d \Gamma(t)}{d t} \sim e^{- \Gamma t} (1 \mp {\rm Im} \lambda_0 \sin
\Delta m t)~~~,
\eeq
where the upper sign refers to $\b$ decays and the lower to $\ob$ decays.
$\Delta m$ is the mass splitting mentioned earlier, and the factor
$\lambda_0$ expresses the interference of decay and mixing amplitudes.  For
$f = J/\psi K_S$, $\lambda_0 = -e^{-2 i \beta}$ to a good approximation, while
for $f = \pi^+ \pi^-$, $\lambda_0 \simeq e^{2 i \alpha}$ only to the extent
that the effect of penguin amplitudes can be neglected in comparison with
the dominant tree contribution.

The time integral of the modulation term is
\beq
\int_0^\infty dt e^{- \Gamma t} \sin \Delta m t = \frac{1}{\Gamma} \frac{x}
{1 + x^2} \le \frac{1}{\Gamma} \cdot \frac{1}{2}~~~,
\eeq
where $x \equiv \Delta m/\Gamma$.  This expression is maximum for $x = 1$,
and 95\% of maximum for the observed value $x \simeq 0.72$.  It has been
fortunate that the $\b$ mixing
amplitude and decay rate are so well matched to one another.

The CDF Collaboration \cite{CDFB} has learned how to ``tag'' neutral $B$
mesons at the time of their production and thus to measure the decay
rate asymmetry in $\b~(\ob) \to J/\psi K_S$.  This asymmetry arises from the
phase $2 \beta$ characterizing the two powers of $V_{td}$ in the $\b$--$\ob$
mixing amplitude.  The tagging methods are of two main types.  ``Opposite-side''
methods rely on the fact that strong interactions always produce $b$ and $\bar
b$ in pairs, so that in order to determine the initial flavor of a decaying $B$
one must find out something about the ``other'' $b$-containing hadron produced
in association with it, either via the charge of the jet containing it or
via the charge of the lepton or kaon it emits when decaying.  ``Same-side''
methods \cite{GNR} utilize the fact that a $\b$ tends to be associated more
frequently with a $\pi^+$, and a $\ob$ with a $\pi^-$, somewhere nearby in
phase space, whether through the dynamics of fragmentation or through the
decays of excited $B$ resonances.

The CDF result is $\sin 2 \beta = 0.79^{+0.41}_{-0.44}$.  An earlier result
from OPAL \cite{OPALB} and a newer result from ALEPH
\cite{ALEPHB}, both utilizing $B$'s produced in the decays of
the $Z^0$, can be combined with the CDF value to obtain $\sin 2 \beta = 0.91
\pm 0.35$, which exceeds zero at the 99\% confidence level \cite{ALEPHB}.
At the $1 \sigma$ lower limit (0.56) this is very close to the lower bound
(0.59) quoted in Table 2.

\subsection{``Self-tagging'' decays and direct CP violation}

An example of direct CP violation can
occur in $\b \to K^+ \pi^-$.  One expects two types of contribution
to this process:  a ``tree'' amplitude governed by the quark subprocess
$\bar b \to \bar u u \bar s$ with CKM factor $V^*_{ub} V_{us}$, and a
``penguin'' amplitude governed by the quark subprocess $\bar b \to \bar s$
with dominant CKM factor $V^*_{tb} V_{ts}$ (since the contribution of the
top quark in the internal loop is dominant).
These contributions are summarized in Table 4.

\begin{table}
\caption{Main amplitudes contributing to $\b \to K^{(*)+} \pi^-$.} 
\begin{center}
\begin{tabular}{c c c c}
Amplitude & Subprocess & CKM factor & Weak phase \\ \hline
Tree      & $\bar b \to \bar u u \bar s$ & $V^*_{ub} V_{us}$ & $\gamma$ \\
Penguin   & $\bar b \to \bar s$ & $V^*_{tb} V_{ts}$ & $\pi$ \\
\end{tabular}
\end{center}
\end{table}

Since the tree and penguin amplitudes have a relative weak phase $\gamma$
(mod $\pi$), one can have $\Gamma(\b \to K^+ \pi^-) \ne \Gamma(\ob \to K^-
\pi^+)$ as long as the strong phases $\delta_T$ and $\delta_P$
are different in the tree and penguin
amplitudes.  However, even if these strong phases do not differ from one
another, the ratios of rates for various charge states of $B \to K \pi$ decays
can provide separate information on the weak phase $\gamma$ \cite{GR,FM,NR} and
the strong phase difference $\delta_T -\delta_P$.

One must first deal with electroweak penguins which were
also relevant for the interpretation of $\epsilon'/\epsilon$.  An early 
suggestion (see the first of Refs.~\cite{GR}) proposed a way to extract
$\gamma$ from the rates for $B^+ \to (\pi^0 K^+, \pi^+ \k, \pi^+ \pi^0)$ and
the charge-conjugate processes.  The amplitudes for the first two processes
(with appropriate factors of $\s$) form a triangle with an amplitude related
to the third process by flavor SU(3) as long as electroweak penguins are
negligible, which they are not \cite{DH}.
It turns out, however \cite{NR}, that the relevant electroweak
penguin's contribution to this process can be calculated, so that sufficiently
precise measurements of the rates for the above processes can indeed yield
useful information on $\gamma$.

The possibility has been raised recently
\cite{NR,GRg,Hou} that the weak phase $\gamma$ may exceed $90^\circ$.  Two
processes whose rates hint at this constraint are $\b \to \pi^+ \pi^-$ and
$\b \to K^{*+} \pi^-$.  The former process has a rate which is somewhat
smaller than expected, while the rate for the latter is larger than expected.

The amplitudes contributing to $\b \to \pi^+ \pi^-$ are summarized in Table 5.
The relative phase of the tree and penguin amplitudes is $\gamma + \beta =
\pi - \alpha$.  The two amplitudes will interfere destructively if the final
strong phase difference is small (as expected from perturbative QCD estimates,
which indeed may be risky), and if $\alpha < \pi/2$.  This would tend to
favor not-too-positive values of $\rho$.  There is some hint that the
interference is indeed destructive.  The observed branching ratio \cite{CLB}
${\cal B}(\b \to \pi^+ \pi^-) = (4.3^{+1.6}_{-1.4} \pm 0.5) \times 10^{-6}$
is less than the value of about $10^{-5}$ which one would estimate \cite{GRg}
from the tree amplitude alone (e.g., using the observed $B \to \pi e \nu_e$
branching ratio and factorization).

\begin{table}
\caption{Main amplitudes contributing to $\b \to \pi^+ \pi^-$.} 
\begin{center}
\begin{tabular}{c c c c}
Amplitude & Subprocess & CKM factor & Weak phase \\ \hline
Tree      & $\bar b \to \bar u u \bar d$ & $V^*_{ub} V_{ud}$ & $\gamma$ \\
Penguin   & $\bar b \to \bar d$ & $V^*_{tb} V_{td}$ & $-\beta$ \\
\end{tabular}
\end{center}
\end{table}

The same types of amplitudes contributing to $\b \to K^+ \pi^-$ also
contribute to $\b \to K^{*+} \pi^-$ (see Table 4).  As in
$\b \to K^+ \pi^-$, the relative phase between the tree and penguin amplitudes
is expected to be $\gamma - \pi$.  One thus expects constructive
interference between the two amplitudes if the strong phase difference
is small and $\gamma > \pi/2$.  Indeed, the branching ratio for $\b \to K^{*+}
\pi^-$ appears to exceed $2 \times 10^{-5}$, while the pure ``penguin''
process $B^+ \to K^+ \phi$ has a branching ratio less than $10^{-5}$.

A global fit to the above two processes and many others (see the second of
Refs.~\cite{Hou}) finds $\gamma = (114^{+24}_{-23})^\circ$, which just grazes
the allowed region quoted in Table 2.  Since the upper bound on $\gamma$ in
Table 2 is set primarily by the lower limit on $B_s$--$\overline{B_s}$ mixing,
such mixing should be visible in experiments of only
slightly greater sensitivity than those performed up to now.

The Tevatron and the LHC
will copiously produce both nonstrange and strange neutral $B$'s,
decaying to $\pi^+ \pi^-$, $K^\pm \pi^\mp$, and $K^+ K^-$ \cite{WurtJesik}.
Each of these channels has particular advantages.

\begin{itemize}

\item The decays
$\b \to K^+ K^-$ and $B_s \to \pi^+ \pi^-$ should be highly
suppressed unless these final states are ``fed'' by rescattering from other
channels \cite{resc}.

\item The decays $\b \to \pi^+ \pi^-$ and $B_s \to K^+ K^-$ can yield $\gamma$
when their time-dependence is measured \cite{RFKK}.  The kinematic peaks
for these two states overlap significantly, so one must either use particle
identification or utilize the vastly different oscillation
frequencies for $\b$--$\ob$ and $B_s$--$\overline{B_s}$ mixing to distinguish
the two final states.

\item A recent proposal for measuring $\gamma$ \cite{bskpi} utilizes the decays
$\b \to K^+ \pi^-$, $B^+ \to \k \pi^+$, $B_s \to K^- \pi^+$, and the
corresponding charge-conjugate processes.  The $\b \to K^+ \pi^-$ and
$B_s \to K^- \pi^+$ peaks are well separated from one another and from
$\b \to \pi^+ \pi^-$ and $B_s \to K^+ K^-$ kinematically \cite{WurtJesik}.

\end{itemize}

The proposal of Ref.~\cite{bskpi} is based on the observation that $B \to K
\pi$ decays involve two types of amplitudes, tree ($T$) and penguin ($P$),
with relative weak phase $\gamma$ and relative strong phase $\delta$.
The decays $B^+ \to \k \pi^+$ are expected to be dominated by the penguin
amplitude (there is no tree contribution except through rescattering from other
final states), so this channel is not expected to display any CP-violating
asymmetries.  One expects $\Gamma(B^+ \to \k \pi^+) = \Gamma(B^- \to \ok
\pi^-)$.  This will provide a check of the assumption that rescattering
effects can be neglected.  A typical amplitude is given by $A(\b \to K^+
\pi^-) = - [P + T e^{i(\gamma + \delta)}]$, where the signs are associated
with phase conventions for states \cite{GHLR}.

We now define
\beq
\left\{ \begin{array}{c} R \\ A_0 \end{array} \right\}
\equiv \frac{\Gamma(\b \to K^+ \pi^-)
\pm \Gamma(\ob \to K^- \pi^+)}{2 \Gamma(B^+ \to \k \pi^+)}~~~,
\eeq
\beq
\left\{ \begin{array}{c} R_s \\ A_s \end{array} \right\}
\equiv \frac{\Gamma(B_s \to K^- \pi^+)
\pm \Gamma(\overline{B_s} \to K^+ \pi^-)}{2 \Gamma(B^+ \to \k \pi^+)}~~~,
\eeq
and $r \equiv T/P$, $\tl \equiv V_{us}/V_{ud}$.  Then one finds
\beq
R = 1 + r^2 + 2 r \cos \delta \cos \gamma~~,~~~
R_s = \tl^2 + \left( \frac{r}{\tl} \right)^2 - 2 r \cos \delta \cos \gamma~~~,
\eeq
\beq
A_0 = - A_s = -2 r \sin \gamma \sin \delta~~~.
\eeq
The sum of $R$ and $R_s$ allows one to determine $r$.  Then using $R$, $r$, and
$A_0$, one can solve for both $\delta$ and $\gamma$.  The prediction $A_s =
- A_0$ serves as a check of the flavor SU(3) assumption which gave these
relations.  An error of $10^\circ$ on $\gamma$ seems feasible with forthcoming
data from Run II of the Tevatron.

The CLEO Collaboration has recently presented some upper limits on CP-violating
asymmetries in $B$ decays to light-quark systems \cite{CLEOCP}, based on 9.66
million events recorded at the $\Upsilon(4S)$.  With asymmetries defined as
\beq
{\cal A}_{CP} \equiv
 \frac{\Gamma(\overline{B} \to \bar f) - \Gamma(B \to f)}
{\Gamma(\overline{B} \to \bar f) + \Gamma(B \to f)}~~~,
\eeq
the results are shown in Table 6.  No statistically significant asymmetries
have been seen yet.  The sensitivity of these results is not yet adequate
to probe the maximum predicted values \cite{comb} $|{\cal A}_{CP}^{K^+ \pi}|
\le 1/3$, but is getting close.

\begin{table}
\caption{CP-violating asymmetries in decays of $B$ mesons to light quarks.}
\begin{center}
\begin{tabular}{c c c}
Mode & Signal events & ${\cal A}_{CP}$ \\ \hline
$K^+ \pi^-$    & $80^{+12}_{-11}$      & $-0.04 \pm 0.16$ \\
$K^+ \pi^0$    & $42.1^{+10.9}_{-9.9}$ & $-0.29 \pm 0.23$ \\
$K_S \pi^+$    & $25.2^{+6.4}_{-5.6}$  & $+0.18 \pm 0.24$ \\
$K^+ \eta'$    & $100^{+13}_{-12}$     & $+0.03 \pm 0.12$ \\
$\omega \pi^+$ & $28.5^{+8.2}_{-7.3}$  & $-0.34 \pm 0.25$ \\ \hline
\end{tabular}
\end{center}
\end{table}

\section{Some future measurements}

The future of the experimental study of CP violation involves a broad program
of experiments with kaons, charmed and $B$ mesons, and neutrinos.  We mention
just a few of the possibilities.

\subsection{Rare kaon decays}

Plans are afoot for measurement of the branching ratio for $K_L \to \pi^0
\nu \bar \nu$ at the required sensitivity (${\cal B} \simeq 3 \times 10^{-3}$).
Experiments are envisioned using both relatively slow kaons at Brookhaven
National Laboratory \cite{K0pio} and faster kaons at the Fermilab Main
Injector \cite{KAMI}.  A Fermilab proposal \cite{CKM} seeks to accumulate
100 events of $K^+ \to \pi^+ \nu \bar \nu$ in order to measure $|V_{td}|$ to
a statistical precision of 5\% and an overall precision of 10\%.

\subsection{Charmed mesons}

Impressive strides have been taken in the measurement of mass differences
and lifetime differences for CP eigenstates of the neutral charmed mesons $D^0$
\cite{CLEOD,FOCUS}.  No significant effects have been seen yet at the level
of a percent or so, but there are tantalizing hints \cite{ChTh}.  It would be
worth while to follow up these possibilities.  Electron-positron colliders,
mentioned below, will devote much of their running time to the study of $B$
mesons, but charmed mesons are accumulated as well in such experiments, and
the samples of them will increase.  Hadronic experiments dedicated to
producing large numbers of $B$'s may also have more to say about mixing,
lifetime differences, and CP violation for charmed mesons.

\subsection{$B$ production in symmetric $e^+ e^-$ collisions}

Although asymmetric $e^+ e^-$ colliders, known as ``B-factories,'' are now
starting to take data at an impressive rate, the CLEO Collaboration at the
symmetric CESR machine has recently celebrated 20 years of $B$ physics,
and is continuing with an active program.  It will be able in the
CLEO-III program to probe charmless $B$ decays down to branching ratios of
$10^{-6}$.  In so doing, it may be able to detect the elusive $\b \to \pi^0
\pi^0$ mode, whose rate will help pin down the penguin amplitude's
contribution and permit a determination of the CKM phase $\alpha$ \cite{GrL}.

Other final states of great interest at this level include $VP$ and $VV$,
where $P,V$ denote light pseudoscalar and vector mesons.  There is a good
chance that direct CP violation may show up in one or more channels if final
state phase differences are sufficiently large.  The detailed study of angular
correlations in $VV$ channels may be able to provide useful information on
strong final state phases.

A useful probe of rescattering effects \cite{resc}, mentioned above, is
the decay $\b \to K^+ K^-$.  This decay is expected to have a branching
ratio of only a few parts in $10^8$ if rescattering is unimportant, but could
be enhanced by more than an order of magnitude in the presence of rescattering
from other channels.

A challenging channel of fundamental importance is $B^+ \to \tau^+ \bar
\nu_\tau$.  The rate for this process will provide information on the
combination $f_B |V_{cb}|$.  Rare decays which have not yet been seen
(such as $B \to X \ell^+ \ell^-$ and $B \to X \nu \bar \nu$) will probe the
effects of new particles in loops.

\subsection{$B$ production in asymmetric $e^+ e^-$ collisions}

The benchmark process for the BaBar and Belle detectors will be the measurement
of $\sin 2 \beta$ in $\b \to J/\psi K_S$.  The PEP-II and KEK-B machines
utilize asymmetric $e^+ e^-$ collisions in order to create a moving reference
frame in which the decays of $\b$ and $\ob$ are separated by a large enough
distance for their separation to be detectable.  (Each travels only an average
distance of 30 $\mu$m in the center of mass.)  This facilitates both flavor
tagging and improvement of signal with respect to background.  These machines
will make possible a host of time-dependent studies in such decays as $B \to
\pi \pi$, $B \to K \pi$, etc., and their impressive luminosities will
eventually add significantly to the world's tally of detected $B$'s.

\subsection{Hadronic $B$ production}

The strange $B$'s cannot be produced at the $\Upsilon(4S)$ which will dominate
the attention of $e^+ e^-$ colliders for some years to come. Hadronic reactions
at high energies will produce copious $b$'s incorporated into
all sorts of hadrons:  nonstrange, strange, and charmed mesons, and baryons.
One looks forward to a measurement of the strange-$B$ mixing
parameter $x_s = \Delta m_s/\Gamma_s$.  The decays of $B_s$ can provide
valueable information on CKM phases and CP violation, as in
$B_s \to K^+ K^-$ \cite{RFKK}.  The width difference of 10--20\% expected
between the CP-even and CP-odd eigenstates of the $B_s$ system \cite{BBD}
should be visible in the next round of experiments.

\subsection{Neutrino studies}

The origin of magnitudes and phases in the CKM matrix is intimately connected
with the origin of the quark masses themselves, whose physics still eludes us.
We will not understand this pattern until we have mapped out a similar pattern
for the leptons, a topic to which many other talks in this Workshop are
devoted.  Our understanding of neutrino masses and mixings
will benefit greatly from forthcoming experiments at the Sudbury Neutrino
Observatory \cite{SNO}, Borexino \cite{Bxo}, K2K \cite{Kam}, and Fermilab
(BooNE and MINOS) \cite{Fnu}, to name a few.

\subsection{The $(\rho,\eta)$ plot in a few years}

\begin{figure}
\centerline{\epsfysize = 3in \epsffile {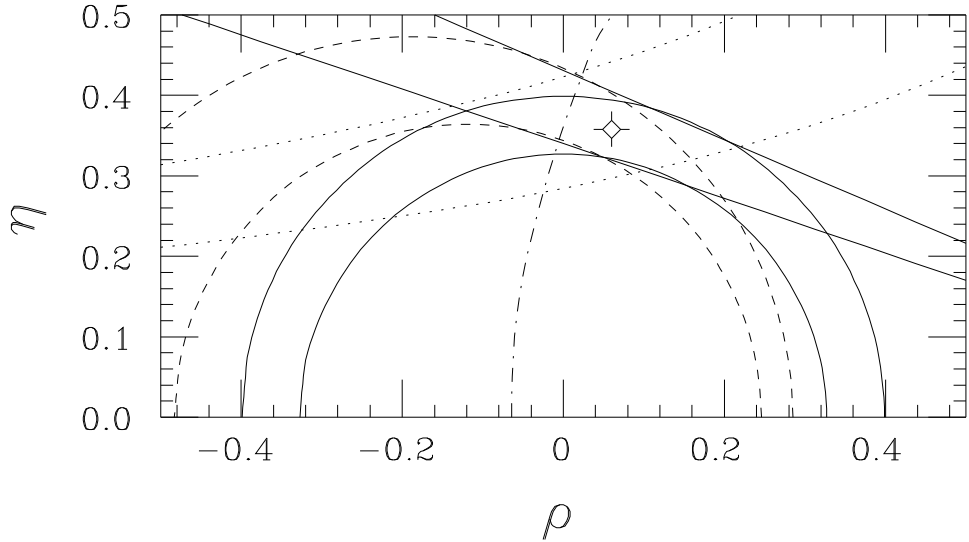}}
\caption{\small Example of a region in the $(\rho,\eta)$ plane that might be
allowed by data in the year 2003.  Constraints are based on the following
assumptions: $|V_{ub}/V_{cb}| = 0.08 \pm 0.008$ (solid semicircles),
$|V_{ub}/V_{td}|=|(\rho-i\eta)/(1-\rho-i\eta)| = 0.362 \pm 0.036$ based on
present data on $B^0$--$\bar B^0$ mixing and a measurement of $B(B^+ \to \tau^+
\nu_\tau)$ to $\pm 20\%$ (dashed semicircles), CP-violating $K$--$\bar K$
mixing as discussed in Sec.~2 but with $V_{cb}$ measured to $\pm 4\%$ (dotted
hyperbolae), the bound $x_s > 20$ for $B_s^0$--$\bar B_s^0$ mixing (to the
right of the dot-dashed semicircle), and measurement of $\sin 2 \beta$ to $\pm
0.059$ (diagonal straight lines).  The plotted point, corresponding to
$(\rho,\eta) = (0.06,0.36)$, lies roughly within the center of the allowed
region.} 
\end{figure}

The $(\rho,\eta)$ plot might appear as shown in Fig.~2 in a few years
\cite{JRlatt,NSF}.  We can look forward either to a reliable determination of
parameters or to the possibility that one or more
experiments give contradictory results, indicating the need for new
physics.  Such new physics most typically shows up in the form of additional
contributions to $\b$--$\ob$ mixing \cite{GLmix}, though it can also show up in
decays \cite{GW}.

\section{Baryon asymmetry}

The ratio of the number of baryons $n_B$ to the number of photons $n_\gamma$ in
the Universe is a few parts in $10^{10}$, much larger than the corresponding
ratio for antibaryons.  Shortly after the discovery of CP violation in neutral
kaon decays, Sakharov proposed in 1967 \cite{Sakh} three ingredients needed to
understand this preponderance of matter over antimatter:  (1) an epoch in which
the Universe was not in thermal equilibrium, (2) an interaction violating
baryon number, and (3) CP (and C) violation.  However, one can't
explain the observed baryon asymmetry merely by means of the CP violation
contained in the CKM matrix.  The effects are too small unless some new physics
is introduced.  Two examples are the following:

\begin{itemize}

\item The concept of supersymmetry, in which each particle of spin $J$ has
a ``superpartner'' of spin $J \pm 1/2$, affords many opportunities for
introducing new CP-violating phases and interactions which could affect
particle-antiparticle mixing \cite{SSBrev}.

\item The presence of neutrino masses at the sub-eV level can signal large
Majorana masses for right-hand neutrinos, exceeding $10^{11}$ GeV \cite{PR}.
Lepton number ($L$) is violated by such masses.  The violation of $L$ can
easily be reprocessed into baryon number ($B$) violation by $B-L$ conserving
interactions at the electroweak scale \cite{LepBar}.  New CP-violating
interactions must then exist at the high mass scale if lepton number is to
be generated there.  It is conceivable that these interactions
are related to CKM phases, but the link will be very indirect \cite{DPF}.
In any case, if this alternative is the correct one, it will be very important
to understand the leptonic analogue of the CKM matrix!

\end{itemize} 

\section{Conclusions}

The CKM theory of CP violation in neutral kaon decays has
passed a crucial test.  The parameter $\epsilon'/\epsilon$ is
nonzero, and has the expected order of magnitude, though exceeding
some theoretical estimates.  Still to come will be
several tests using $B$ mesons, including the observation of a difference in
rates between $\b \to J/\psi K_S$ and $\ob \to J/\psi K_S$.  There will be
more progress in ``tagging'' neutral $B$'s, and we
can look forward to rich information from measurements of decay
rates of charged and neutral $B$'s into a variety of final states.

I see two possibilities for our understanding of CP violation in the next few
years. (1) If $B$ decays do not provide a consistent set of CKM phases, we will
be led to examine other sources of CP violation.
Most of these, in contrast to the CKM theory, predict neutron
and electron dipole moments very close to their present experimental upper
limits. (2) If, on the other hand, the CKM picture still hangs together after
a few years, attention should naturally shift to the next ``layer of the
onion'': the origin of the CKM phases (and the associated quark and
lepton masses).  It is probably time to start anticipating this possibility,
given the resilience of the CKM picture since it was first proposed nearly
30 years ago.

\section{Acknowledgements}

It is a pleasure to thank Maria Eugenia and Jose Nieves for their wonderful
hospitality in San Juan.  This work was supported in part by the United States
Department of Energy under Grant No.\ DE FG02 90ER40560.

\def \ajp#1#2#3{Am.\ J. Phys.\ {\bf#1}, #2 (#3)}
\def \apny#1#2#3{Ann.\ Phys.\ (N.Y.) {\bf#1}, #2 (#3)}
\def \app#1#2#3{Acta Phys.\ Polonica {\bf#1}, #2 (#3)}
\def \arnps#1#2#3{Ann.\ Rev.\ Nucl.\ Part.\ Sci.\ {\bf#1}, #2 (#3)}
\def \cmts#1#2#3{Comments on Nucl.\ Part.\ Phys.\ {\bf#1}, #2 (#3)}
\def \cn{Collaboration}
\def \cp89{{\it CP Violation,} edited by C. Jarlskog (World Scientific,
Singapore, 1989)}
\def \epjc#1#2#3{Eur.~Phys.~J.~C {\bf#1}, #2 (#3)}
\def \f79{{\it Proceedings of the 1979 International Symposium on Lepton and
Photon Interactions at High Energies,} Fermilab, August 23-29, 1979, ed. by
T. B. W. Kirk and H. D. I. Abarbanel (Fermi National Accelerator Laboratory,
Batavia, IL, 1979}
\def \hb87{{\it Proceeding of the 1987 International Symposium on Lepton and
Photon Interactions at High Energies,} Hamburg, 1987, ed. by W. Bartel
and R. R\"uckl (Nucl. Phys. B, Proc. Suppl., vol. 3) (North-Holland,
Amsterdam, 1988)}
\def \ib{{\it ibid.}~}
\def \ibj#1#2#3{~{\bf#1}, #2 (#3)}
\def \ichep72{{\it Proceedings of the XVI International Conference on High
Energy Physics}, Chicago and Batavia, Illinois, Sept. 6 -- 13, 1972,
edited by J. D. Jackson, A. Roberts, and R. Donaldson (Fermilab, Batavia,
IL, 1972)}
\def \ijmpa#1#2#3{Int. J. Mod. Phys. A {\bf#1}, #2 (#3)}
\def \ite{{\it et al.}}
\def \jhep#1#2#3{JHEP {\bf#1}, #2 (#3)}
\def \jpb#1#2#3{J.~Phys.~B~{\bf#1}, #2 (#3)}
\def \lg{{\it Proceedings of the XIXth International Symposium on
Lepton and Photon Interactions,} Stanford, California, August 9--14 1999,
edited by J. Jaros and M. Peskin (World Scientific, Singapore, 2000)}
\def \lkl87{{\it Selected Topics in Electroweak Interactions} (Proceedings of
the Second Lake Louise Institute on New Frontiers in Particle Physics, 15 --
21 February, 1987), edited by J. M. Cameron \ite~(World Scientific, Singapore,
1987)}
\def \kdvs#1#2#3{{Kong.~Danske Vid.~Selsk., Matt-fys.~Medd.} {\bf #1}, No.~#2
(#3)}
\def \ky85{{\it Proceedings of the International Symposium on Lepton and
Photon Interactions at High Energy,} Kyoto, Aug.~19-24, 1985, edited by M.
Konuma and K. Takahashi (Kyoto Univ., Kyoto, 1985)}
\def \mpla#1#2#3{Mod. Phys. Lett. A {\bf#1}, #2 (#3)}
\def \nat#1#2#3{Nature {\bf#1}, #2 (#3)}
\def \nc#1#2#3{Nuovo Cim. {\bf#1}, #2 (#3)}
\def \np#1#2#3{Nucl. Phys. {\bf#1}, #2 (#3)}
\def \PDG{Particle Data Group, L. Montanet \ite, \prd{50}{1174}{1994}}
\def \pisma#1#2#3#4{Pis'ma Zh. Eksp. Teor. Fiz. {\bf#1}, #2 (#3) [JETP Lett.
{\bf#1}, #4 (#3)]}
\def \pl#1#2#3{Phys. Lett. {\bf#1}, #2 (#3)}
\def \pla#1#2#3{Phys. Lett. A {\bf#1}, #2 (#3)}
\def \plb#1#2#3{Phys. Lett. B {\bf#1}, #2 (#3)}
\def \pr#1#2#3{Phys. Rev. {\bf#1}, #2 (#3)}
\def \prc#1#2#3{Phys. Rev. C {\bf#1}, #2 (#3)}
\def \prd#1#2#3{Phys. Rev. D {\bf#1}, #2 (#3)}
\def \prl#1#2#3{Phys. Rev. Lett. {\bf#1}, #2 (#3)}
\def \prp#1#2#3{Phys. Rep. {\bf#1}, #2 (#3)}
\def \ptp#1#2#3{Prog. Theor. Phys. {\bf#1}, #2 (#3)}
\def \rmp#1#2#3{Rev. Mod. Phys. {\bf#1}, #2 (#3)}
\def \rp#1{~~~~~\ldots\ldots{\rm rp~}{#1}~~~~~}
\def \si90{25th International Conference on High Energy Physics, Singapore,
Aug. 2-8, 1990}
\def \slc87{{\it Proceedings of the Salt Lake City Meeting} (Division of
Particles and Fields, American Physical Society, Salt Lake City, Utah, 1987),
ed. by C. DeTar and J. S. Ball (World Scientific, Singapore, 1987)}
\def \slac89{{\it Proceedings of the XIVth International Symposium on
Lepton and Photon Interactions,} Stanford, California, 1989, edited by M.
Riordan (World Scientific, Singapore, 1990)}
\def \smass82{{\it Proceedings of the 1982 DPF Summer Study on Elementary
Particle Physics and Future Facilities}, Snowmass, Colorado, edited by R.
Donaldson, R. Gustafson, and F. Paige (World Scientific, Singapore, 1982)}
\def \smass90{{\it Research Directions for the Decade} (Proceedings of the
1990 Summer Study on High Energy Physics, June 25--July 13, Snowmass, Colorado),
edited by E. L. Berger (World Scientific, Singapore, 1992)}
\def \tasi{{\it Testing the Standard Model} (Proceedings of the 1990
Theoretical Advanced Study Institute in Elementary Particle Physics, Boulder,
Colorado, 3--27 June, 1990), edited by M. Cveti\v{c} and P. Langacker
(World Scientific, Singapore, 1991)}
\def \yaf#1#2#3#4{Yad. Fiz. {\bf#1}, #2 (#3) [Sov. J. Nucl. Phys. {\bf #1},
#4 (#3)]}
\def \zhetf#1#2#3#4#5#6{Zh. Eksp. Teor. Fiz. {\bf #1}, #2 (#3) [Sov. Phys. -
JETP {\bf #4}, #5 (#6)]}
\def \zpc#1#2#3{Zeit. Phys. C {\bf#1}, #2 (#3)}
\def \zpd#1#2#3{Zeit. Phys. D {\bf#1}, #2 (#3)}

\end{document}